\pdfoutput=1

\documentclass[journal]{IEEEtran}
\ifCLASSINFOpdf
   \usepackage[pdftex]{graphicx}
   \graphicspath{{../pdf/}{../jpeg/}}
   \DeclareGraphicsExtensions{.pdf,.jpeg,.png}
\else
\fi
\usepackage{latexsym}
\usepackage[cmex10]{amsmath}
\newcommand{\str}[1]{#1}	

\newcommand{\MYfooter}{\smash{\scriptsize
\hfil\parbox[t][\height][t]{\textwidth}{\centering
~\\
~}\hfil\hbox{}}}

\newcommand{\MYarxivheader}{\smash{\scriptsize
\hfil\parbox[t][\height][t]{\textwidth}{\centering
This work has been submitted to the IEEE for possible publication. Copyright may be transferred without notice, after which this version may no longer be available.
}\hfil\hbox{}}}

\makeatletter

\def\ps@headings{%
\def\@oddhead{\mbox{}\scriptsize\rightmark \hfil \thepage}
\def\@evenhead{\scriptsize\thepage \hfil \leftmark\mbox{}}
\def\@oddfoot{\MYfooter}%
\def\@evenfoot{\MYfooter}}

\def\ps@IEEEtitlepagestyle{%
\def\@oddhead{\MYarxivheader}%
\def\@evenhead{\scriptsize\thepage \hfil \leftmark\mbox{}}%
\def\@oddfoot{\scriptsize\thepage \hfil \leftmark\mbox{}}%
\def\@evenfoot{\scriptsize\thepage \hfil \leftmark\mbox{}}}
\def\@oddfoot{\MYfooter}%
\def\@evenfoot{\MYfooter}

\makeatother

\begin{document}
%
\title{Static Non-linearity in Graphene Field Effect Transistors}
%
%
%

\author{Saul~Rodriguez,~\IEEEmembership{Member,~IEEE,}
	Anderson~Smith,~\IEEEmembership{Student Member,~IEEE,}       
	Sam~Vaziri,~\IEEEmembership{Student Member,~IEEE,}      
	Mikael~Ostling,~\IEEEmembership{Fellow,~IEEE,}
        Max~C.~Lemme,~\IEEEmembership{Senior Member,~IEEE,}
       and~Ana~Rusu,~\IEEEmembership{Member,~IEEE}

\thanks{Manuscript received March 12, 2014. Support from the European Commission through a STREP project (GRADE, No. 317839), an ERC Advanced Investigator Grant (OSIRIS, No. 228229), and an ERC Starting Grant (InteGraDe, No. 307311) as well as the German Research Foundation (DFG, LE 2440/1-1) is gratefully acknowledged.}
\thanks{S. Rodriguez, A. Smith, S. Vaziri, M. Ostling, and A. Rusu are with the KTH Royal Institute of Technology, School of ICT, Kista, Sweden (email: saul@kth.se; andsmi@kth.se; vaziri@kth.se; ostling@kth.se; arusu@kth.se)}
\thanks{M. C. Lemme is with the University of Siegen, Graphene-based Nanotechnology, Germany (email: max.lemme@uni-siegen.de)}}

\maketitle


\begin{abstract}
The static linearity performance metrics of the GFET transconductor are studied and modeled. Closed expressions are proposed for second and third order harmonic distortion ($HD_2$, $HD_3$), second and third order intermodulation  distortion ($\Delta IM_2$, $\Delta IM_3$), and second and third order intercept points ($A_{IIP2}$, $A_{IIP3}$). The expressions are validated through large-signal simulations using a GFET VerilogA analytical model and a commercial circuit simulator. The proposed expressions can be used during circuit design in order to predict the GFET biasing conditions at which linearity requirements are met.
\end{abstract}


%
\IEEEpeerreviewmaketitle


\section{Introduction}
%
%
%
%
\IEEEPARstart{G}{raphene} based field effect transistors (GFETs) are nowadays considered as a technology option for future high-speed RF circuits and systems. Performance metrics such as high intrinsic transit frequency $f_T$ and transconductance gain $g_m$ have shown very competitive performance when compared to similarly sized devices of other technologies \cite{Lin2011a}, \cite{Wu2011a}, \cite{Liao2010a}. 
Another key performance  metric that is important in RF systems and that must be carefully characterized in GFETs is the linearity.

RF circuits must process weak signals in the presence of strong interference; therefore, they must exhibit high linearity performance. Non-linearities in the RF circuits are the source of several undesirable effects such as harmonic distortion, gain compression, intermodulation, cross-modulation, AM/PM conversion, DC offsets, etc \cite{Razavi2012}. Therefore, it is of paramount importance to study and characterize the intrinsic non-linearities of the GFET devices. Initial measurements of linearity of GFETs have shown that is possible to achieve good linearity performance \cite{Jenkins2013}. However, an analytic study that allows to identify the GFET linearity under different design parameters and biasing conditions is required.

\begin{figure}[t!]
\centering
\includegraphics[width=6.5cm]{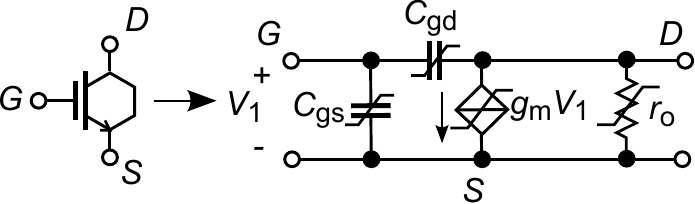}
\caption{Non-linear small-signal representation of GFET devices.}
\label{FIG:ID}
\end{figure}

Fig.~\ref{FIG:ID} shows the small-signal representation of the GFET. The small-signal representation assumes that the components  are linear. \str{However, the capacitances $C_{gs}$, $C_{gd}$, transconductance $g_m$, and output resistance $r_{o}$ are in fact non-linear components.} 
The capacitors are storage elements which exhibit memory effects. Their non-linearity is dynamic and as can be expected, its effect is more visible at high frequencies. 
The parameters $g_m$ and $r_o$ have a memoryless behavior and therefore their non-linearities are static. The effect of the non-linearity of $r_o$ depends mainly on the loading conditions at the drain of the device. In most common cases, there are impedances smaller than  $r_o$ loading the drain. Therefore, the impact of the non-linearities of $r_o$ is generally reduced. 
The non-linearity of $g_m$, on the other hand,  can not be reduced unless negative feedback is intentionally applied at the expense of noise and gain degradation. The non-linearity of $g_m$, in fact, places a fundamental limit on the total non-linearity of the device. This paper introduces closed expressions for the second and third order static non-linearity of $g_m$ which are the main concerns in RF applications. 
\section{Transconductance Non-linearity }
The static non-linearity at the drain current can be approximated by using a Taylor expansion polynomial:
\begin{align}
I_{D} = a_1v_{in}(t)+a_2v_{in}^2(t)+a_3v_{in}^3(t)+...+a_nv_{in}^n(t)
\label{EQ:POL}
\end{align}
where $I_D$ is the drain current, and $v_{in}$ is a input voltage signal at the gate. The first three coefficients $(a_1,a_2,a_3)$ generally dominate the non-linearities for small signals. Higher order terms appear when the input signals are large enough so that the device leaves the saturation region and stops acting as an amplifier. This condition is known as clipping and can be avoided by ensuring proper signal levels at the input. The first terms can be found by differentiating (\ref{EQ:POL}) and solving for the unknown coefficients:
\begin{align}
a_1 = \left. \frac{\delta I_{D}}{\delta v_{in}} \right |_{v= 0}
a_2 = \left. \frac{1}{2} \frac{\delta^2 I_{D}}{\delta v_{in}^2} \right |_{v= 0}
a_3 = \left. \frac{1}{6} \frac{\delta^3 I_{D}}{\delta v_{in}^3} \right |_{v=0}
\end{align}

The drain current in GFETs can be expressed as \cite{Rodriguez2014} :
\begin{align}
I_D = \frac{\mu W C_{TOP} \left ( V_{eff} - {V_{DSi}}/{2} \right )}{{L}/{V_{DSi}} + \frac{\mu }{\omega }\sqrt{{\pi C_{TOP}}/{e}}\sqrt{V_{eff}-V_{DSi}/2}}
\label{EQ:ID1}
\end{align}

where $\mu$ is the mobility, $W$ the transistor width, $L$ the transistor length,  $C_{TOP}$ is the top oxide capacitance density, 
$e$ the elementary charge, $\omega$ is obtained from the surface phonon energy of the substrate $\hbar\omega $, and $V_ {eff} = V_{GSi} + V_{TH,0}$. The zero bias threshold voltage $V_{TH,0} = eN_f/C_{TOP}$ accounts for the shift in the Dirac point due to the doping level $N_f$. Equation (\ref{EQ:ID1}) is valid for the first triode  and saturation/negative resistance regions when $V_{eff} > V_{DSi}/2$ and ${\Delta ^2}/{\pi \hbar^2 {v_f}^2 } \ll  \pi{ \left | Q_{NET,AV}\right |}/{e}$. 
By performing several substitutions,  (\ref{EQ:ID1}) can be expressed as:
\begin{align}
I_D = \frac{A x }{{L}/{V_{DSi}} + B\sqrt{x}}
\label{EQ:ID2}
\end{align}
where $A=\mu W C_{TOP}$, $B = \frac{\mu }{\omega }\sqrt{{\pi C_{TOP}}/{e}}$, and $x~=~V_{eff}~-~V_{DSi}/2$.   The coefficients $a_1 - a_3$ are found by taking derivatives of (\ref{EQ:ID2}) with respect to the variable $x$ ($\delta x = \delta V_{eff} = \delta V_{GSi}$), and replacing the results in (2). Accordingly, the polynomial coefficients are:
\begin{align}
a_1 = \frac{A V_{DSi}\left ( 2L+B V_{DSi} \sqrt{x} \right )}{2\left ( L + B V_{DSi}\sqrt{x} \right )^2}
\label{EQ:a1}
\end{align}
\begin{align}
a_2 = -\frac{A V_{DSi} \left ( B^2 V_{DSi}^2 x + 3 B L V_{DSi} \sqrt{x}  \right )}{8x\left ( L+BV_{DSi} \sqrt{x} \right )^3}
\label{EQ:a2}
\end{align}
 \begin{align}
a_3 = \frac{ABV_{DSi}^2\left ( L^2+B^2V_{DSi}^2x + 4BLV_{DSi} \sqrt{x} \right )}
{16 x^{3/2} \left ( L + B V_{DSi} \sqrt{x} \right )^4}
\label{EQ:a3}
\end{align}
Once (\ref{EQ:a1}), (\ref{EQ:a2}), and (\ref{EQ:a3}) are substituted in (\ref{EQ:POL}), the resulting polynomial depends only on technology parameters and biasing voltages, and  can be used to calculate the harmonic and intermodulation distortion of GFET devices.
\subsection{Second and Third Harmonic Distortion}
The second and third order harmonic distortion levels ($HD_2$ and $HD_3$) produced by a single input tone $v_m \cos{(\omega_1 t)}$ are found by calculating the ratio of the output currents at the harmonics frequencies ($2\omega_1$, $3\omega_1$) to the output current at the fundamental frequency $\omega1$. The second harmonic distortion is given by:
\begin{align}
HD_2 = \frac{1}{2} \left | \frac{a_2}{a_1} \right |v_m
\label{EQ:d2}
\end{align}
\begin{align}
HD_2 = \frac{v_mL^2}{4x\left ( L+BV_{DSi} \sqrt{x} \right )\left ( 2L + BV_{DSi} \sqrt{x} \right )} - \frac{v_m}{8x}
\end{align}
The third harmonic distortion is:
\begin{align}
HD_3 = \frac{1}{4} \left | \frac{a_3}{a_1} \right |v_m^2
\label{EQ:d3}
\end{align}
\begin{equation}
\begin{split}
&HD_3 = \left ( \frac{BV_{DSi}\left ( xB^2V_{DSi}^2+L^2 \right )}{32x^{3/2}}+\frac{B^2LV_{DSi}^2}{8x} \right )\bigg/ \\
&\left [ \left ( L+BV_{DSi} \sqrt{x} \right )^2 \left ( 2L + BV_{DSi}\sqrt{x} \right ) \right ] \times v_m^2
\end{split}
\end{equation}
\subsection{Intermodulation Distortion}
The 	second order intermodulation distortion ($\Delta IM_2$) produced by two input tones  $v_m \cos{(\omega_1 t)}$ and  $v_m \cos{(\omega_2 t)}$ is found by calculating the ratio of  the output intermodulation product at $\omega_1 \pm \omega_2$  to the output at any of the fundamental frequencies ($\omega_1$,$\omega_2$):
\begin{align}
\Delta IM_2 = \left | \frac{a_2}{a_1} \right |v_m
\label{EQ:IM2}
\end{align}
\begin{equation}
\begin{split}
&\Delta IM_2 = \\
& \frac{v_mL^2}{2x\left ( L+BV_{DSi} \sqrt{x} \right )\left ( 2L + BV_{DSi} \sqrt{x} \right )} - \frac{v_m}{4x}
\end{split}
\end{equation}

The 	third order intermodulation distortion ($\Delta IM_3$) produced by two input tones  $v_m \cos{(\omega_1 t)}$ and  $v_m \cos{(\omega_2 t)}$ is found by calculating the ratio of  the output intermodulation product at ($2\omega_1 - \omega_2$, $2\omega_2-\omega_1$)  to the output at any of the fundamental frequencies ($\omega_1$,$\omega_2$):
\begin{align}
\Delta IM_3 = \frac{3}{4} \left | \frac{a_3}{a_1} \right |v_m^2
\label{EQ:d3}
\end{align}
\begin{equation}
\begin{split}
&\Delta IM_3 = \left ( \frac{BV_{DSi}\left ( xB^2V_{DSi}^2+L^2 \right )}{32x^{3/2}}+\frac{B^2LV_{DSi}^2}{8x} \right ) \\
& \bigg/ \left [ \left ( L+BV_{DSi} \sqrt{x} \right )^2 \left ( 2L + BV_{DSi}\sqrt{x} \right ) \right ] \times 3  v_m^2
\end{split}
\end{equation}
\subsection{Second and Third Order Intercept Points}

While the $HD_2$, $HD_3$, $\Delta IM_2$, and $\Delta IM_3$ allow quick estimations of the distortion levels for a given input voltage $v_m$, they are not suitable as figures of merit for comparing the linearity performance.  These comparisons are normally done using the so-called second and third order intercept points ($A_{IIP2}$, $A_{IIP3}$). 
The $IIP_2$ is defined as:
\begin{align}
A_{IIP2} = \left | \frac{a_1}{a_2} \right |
\label{EQ:IIP2}
\end{align}
\begin{align}
A_{IIP2}=4x+\frac{8L^2\sqrt{x}}{B V_{DSi}\left ( 3L + B V_{DSi} \sqrt{x} \right )}
\end{align}
The $A_{IIP3}$ is defined as:
\begin{align}
A_{IIP3} = \sqrt{ \frac{4}{3} \left | \frac{a_1}{a_3} \right |}
\label{EQ:IIP3}
\end{align}
\begin{equation}
\begin{split}
& A_{IIP3} = \\
&  \sqrt{\frac{32x^{3/2}\left ( L + B V_{DSi}\sqrt{x} \right )^2 \left ( 2L +B V_{DSi} \sqrt{x} \right )}{3BV_{DSi} \left ( L^2+B^2V_{DSi}^2x + 4BLV_{DSi} \sqrt{x}  \right )}}
\end{split}
\end{equation}
\subsection{Validation of Linearity Expressions}
\str{The best way to validate the proposed linearity expressions is to compare the linearity metrics calculated by using these expressions with linearity measurements on GFET devices.  When such measurements are unavailable, an alternative approach is to estimate these linearity metrics by performing large-signal time-domain transient simulations using transistor models. A caveat of this approach is that the accuracy of these estimations depends on how accurate the transistor model is.  In this work, linearity metrics are estimated by performing large-signal simulation using a commercial circuit simulator (Cadence Spectre) and the GFET Verilog-A model from \cite{Fregonese2013}. This model has been successfully used and verified by using measurements of differently sized GFETs fabricated by different groups.  }
The test device is a 440 nm length, 1 $\mu$m width GFET from \cite{Meric2010a}.  
The model parameters are  $N_f \approx 0$ ($V_{TH,0} \approx 0$ V), $\varepsilon_r = 3.5$, $T_{OX} = 8.5$ nm ($C_{TOP} = 3.6 \times 10^{-3} \text{F}/\text{m}^2$), $\mu = 7000$~$\text{cm}^2\text{V}^{-1}\text{s}^{-1}$, $\Delta = 66.8$ meV, and   $\hbar\omega = 56~\text{meV}$. The GFET is biased at $V_{DS}$ = 0.4 V and $V_{GS}$ = 1.5 V. Single and two-tone periodic-state simulations are performed in order to obtain the harmonic content of the drain current. The amplitude of the input signals $v_m$ is set at   20~mV. The frequency of the input signal for the single-tone test is 1~kHz while the frequencies for the two-tone test are 10~kHz and 11~kHz.

\begin{figure}[t!]
\centering
\includegraphics[width=6cm]{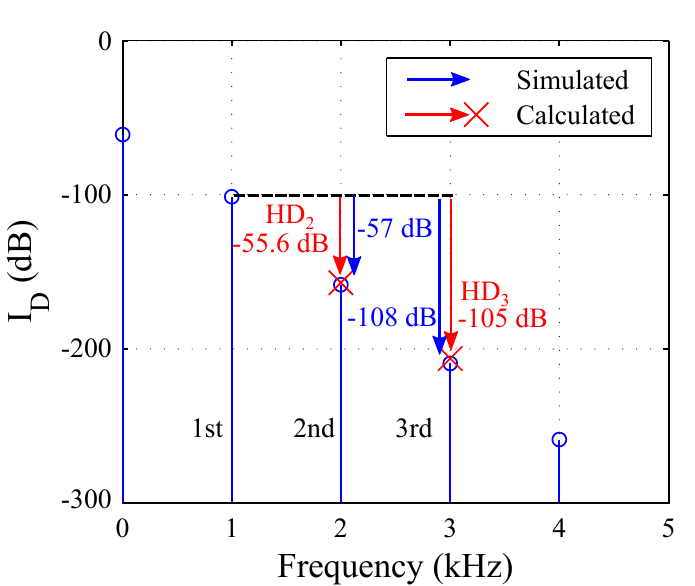}
\caption{Single-tone harmonic distortion test.}
\label{FIG:HAR}
\end{figure}

\begin{figure}[t!]
\centering
\includegraphics[width=6cm]{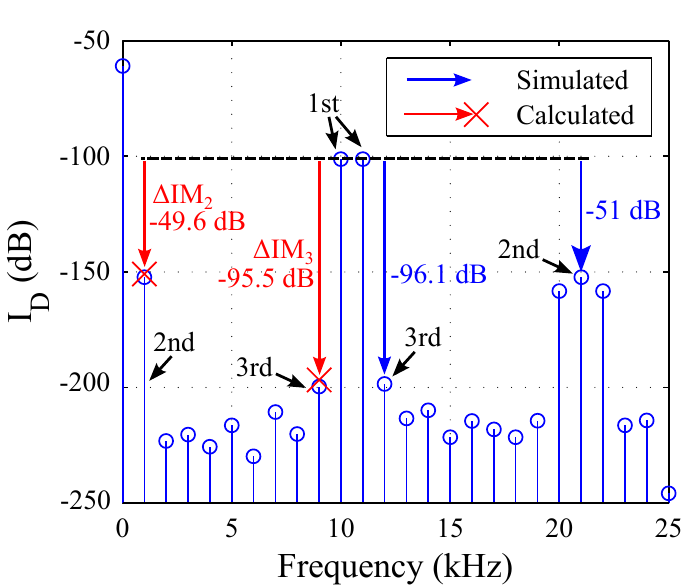}
\caption{Two-tone intermodulation test.}
\label{FIG:IIP}
\end{figure}

Fig. \ref{FIG:HAR} and Fig. \ref{FIG:IIP}  show the simulated spectrum content of the drain current for the single-tone  and two-tone tests, and the simulated and calculated distortion levels. It can be seen that the calculated values of $HD_2$, $HD_3$, $\Delta IM_2$, and $\Delta IM_3$ predict very accurately the simulated distortion levels. The simulated second and third order intercept points are found by using the expressions $A_{IIP2,sim}(dBV) = 20log_{10}(v_m)+|IM_{2,sim}|(dB)$, and $A_{IIP3,sim}(dBV) = 20log_{10}(v_m)+|IM_{3,sim}|/2(dB)$. Table~I shows a summary of the calculated  and simulated linearity metrics. \str{In addition, the table shows simulated linearity metrics for an equally sized NMOS device (440 nm lenght, 1 $\mu$m width) which was biased at the same $V_{DS}$ and $V_{GS}$ voltages as the test GFET. The NMOS device belongs to a 0.15$\mu$m CMOS commercial process and these metrics were extracted using the same large-signal simulations in Cadence Spectre. It can be appreciated that the GFET device outperforms the MOS device under similar conditions. This is a very encouraging result that shows the potential of GFET devices for highly-linear RF circuits.}

\section{Conclusions}
This paper has presented analytical expressions for the  GFET transconductance non-linearity, which is the main source of  distortion. The proposed expressions can be efficiently used to predict the linearity performance metrics of GFETs under different technology parameter values and biasing conditions. These expressions enable  to perform comparisons of linearity performance of GFETs with other transistor technologies. \str{As an example, a linearity comparison between a GFET and a CMOS device was performed. The comparison shows that the GFET outperforms its similarly sized  CMOS  counterpart. Therefore, GFET devices can potentially be used to design highly-linear RF circuits.}

\begin{table}[!t]
\renewcommand{\arraystretch}{1.5}
\caption{Simulated and Calculated Linearity}
\centering
\begin{tabular}{c c c c}
\hline
\bfseries Name & \bfseries Calculated & \bfseries Simulated  & \bfseries Simulated\\
\hline\hline
Technology & GFET & GFET & CMOS \\
$HD_2$ &   -55.6 dB   & -57 dB & -45.7 dB\\
$HD_3$ &   -105 dB  &  -108 dB & -96.7 dB\\
$\Delta IM_2$ & -49.6 dB  & -51 dB & -39.7 dB\\
$\Delta IM_3$ & -95.5 dB & -96.1 dB & -87.2 dB\\
$A_{IIP2}$ & 15.7 dBV  &  17 dBV & 5.7 dBV\\
$A_{IIP3}$ & 13.8 dBV & 14 dBV & 9.6 dBV\\
\hline

\end{tabular}
\label{TABLE:RES}
\end{table}

\ifCLASSOPTIONcaptionsoff
  \newpage
\fi



\bibliographystyle{IEEEtran}
\bibliography{GRAPHENE_CIRCUITS}
\end{document}